\documentclass[preprint,12pt,authoryear]{elsarticle}

\usepackage{amssymb}
\usepackage{amsmath}
\usepackage{csquotes}
\usepackage{graphicx}
\usepackage{multirow}
\usepackage{amsfonts}%
\usepackage{amsthm}%
\usepackage{mathrsfs}%
\usepackage{subcaption}
\usepackage{xcolor}%
\usepackage{textcomp}%
\usepackage{manyfoot}%
\usepackage{booktabs}%
\usepackage{algorithm}%
\usepackage{algorithmicx}%
\usepackage{algpseudocode}%
\usepackage{listings}%
\usepackage{dirtytalk}
\usepackage{csquotes}
\usepackage{hyperref}
\usepackage{tcolorbox}
\journal{Engineering Applications of Artificial Intelligence}

\begin{document}

\begin{frontmatter}



\title{Visual Grounding Methods for Efficient Interaction with Desktop Graphical User Interfaces}


%\author{} %% Author name

\author[1]{El Hassane Ettifouri}
\ead{novylab.research@novelis.io}

\author[1]{Jessica López Espejel}
\ead{jlopezespejel@novelis.io}

\author[1]{Laura Minkova}
\ead{lminkova@novelis.io}

\author[1]{Tassnim Dardouri}
%\email{tdardouri@novelis.io}

\author[1]{Walid Dahhane}
\ead{wdahhane@novelis.io}



\affiliation[1]{
    organization={Research and Innovation Lab, Novelis}, city={Paris}, 
    postcode={75012}, 
    state={Île-de-France}, 
    country={France}
    }

%% Abstract
\begin{abstract}
    Most visual grounding solutions primarily focus on realistic images. However, applications involving synthetic images, such as Graphical User Interfaces (GUIs), remain limited. This restricts the development of autonomous computer vision-powered artificial intelligence (AI) agents for automatic application interaction. Enabling AI to effectively understand and interact with GUIs is crucial to advancing automation in software testing, accessibility, and human-computer interaction. In this work, we explore Instruction Visual Grounding (IVG), a multi-modal approach to object identification within a GUI. More precisely, given a natural language instruction and a GUI screen, IVG locates the coordinates of the element on the screen where the instruction should be executed. We propose two main methods: (1) IVGocr, which combines a Large Language Model (LLM), an object detection model, and an Optical Character Recognition (OCR) module; and (2) IVGdirect, which uses a multimodal architecture for end-to-end grounding. For each method, we introduce a dedicated dataset. In addition, we propose the Central Point Validation (CPV) metric, a relaxed variant of the classical Central Proximity Score (CPS) metric. Our final test dataset is publicly released to support future research.
\end{abstract}

\begin{graphicalabstract}
    \begin{center}
        \textbf{IVGocr architecture}
        \includegraphics[width=0.85\textwidth]{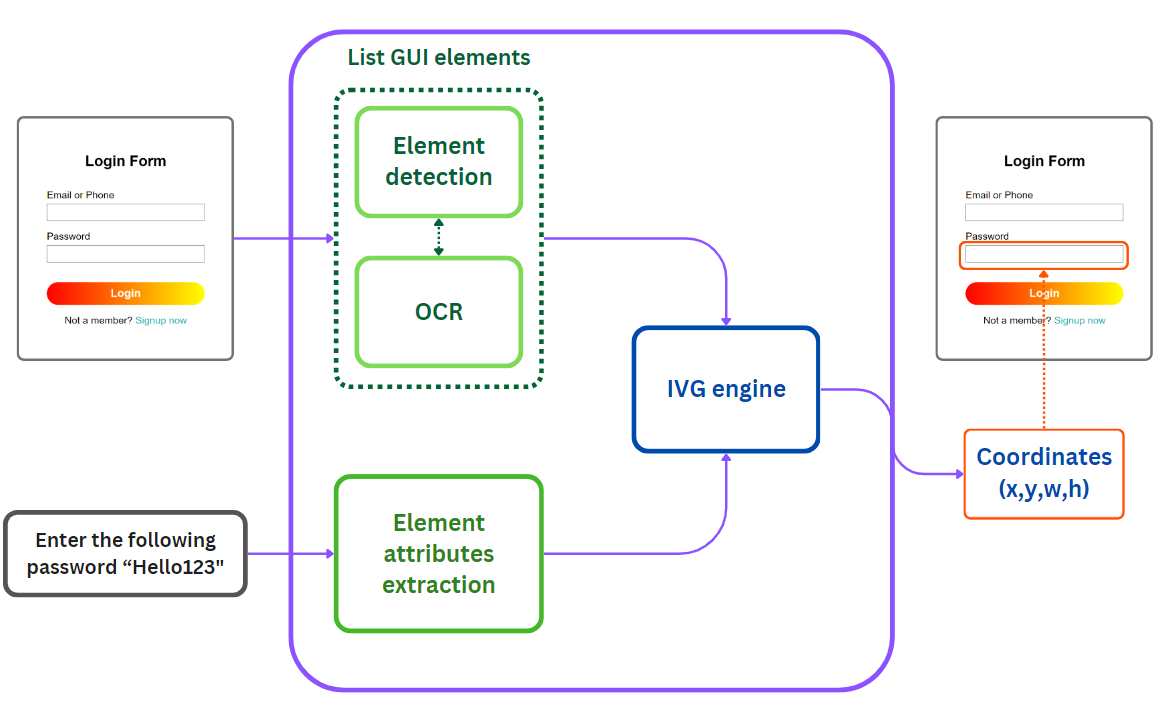}
    \end{center}
    
    \begin{center}
        \textbf{IVGdirect architecture}
        \includegraphics[width=0.85\textwidth]{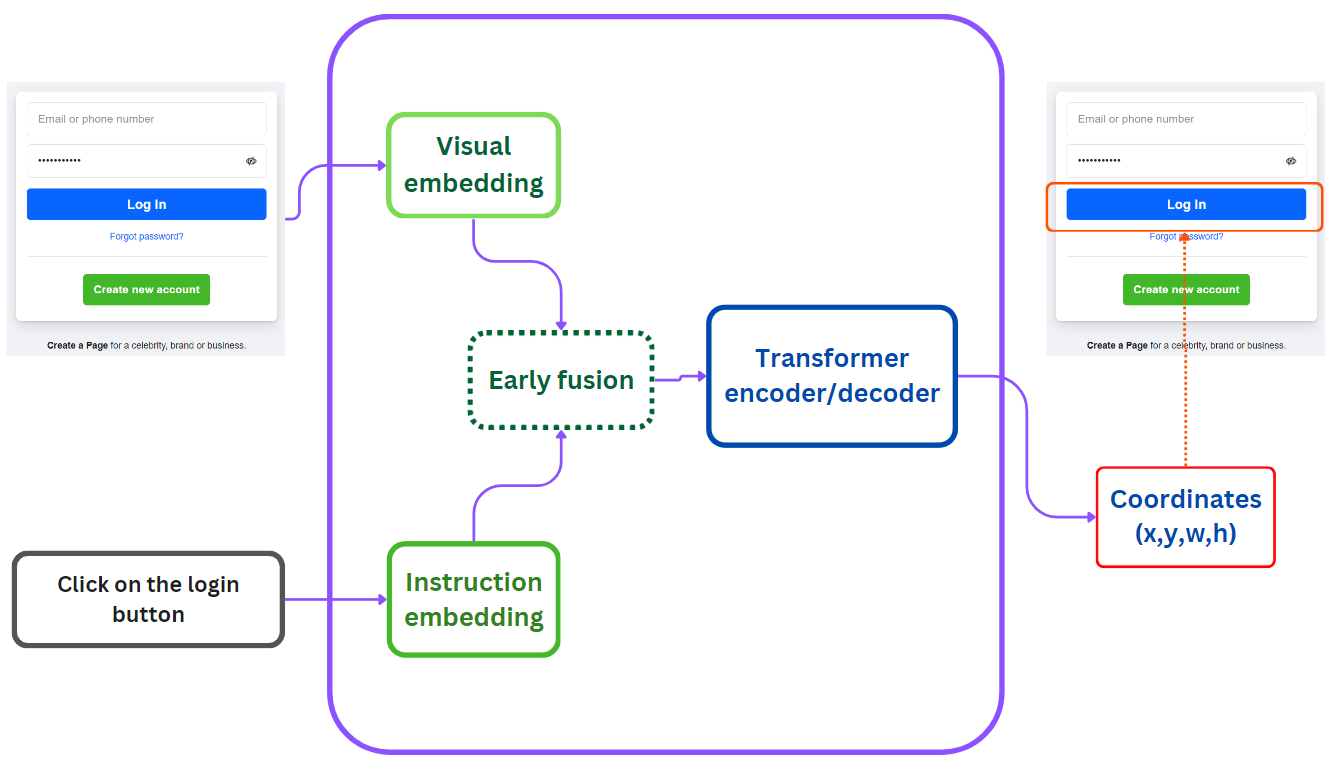}
    \end{center}
\end{graphicalabstract}

\begin{highlights}
    \item 
    We introduce IVGocr (Instruction Visual Grounding with OCR), a new dataset and method that uses a combination of ChatGPT~\citep{openai2024gpt4}, an object detection model (YOLOv8~\citep{reis2023realtime}), and an OCR tool to help understand and follow instructions on graphical user interfaces (GUIs).
    \item
    We present IVGdirect (Instruction Visual Grounding Direct), a simpler method and dataset that trains a vision-language model—based on previous work by~\cite{yan2023universal} directly understand instructions in GUI environments.
    \item
    We introduce a new evaluation metric called Central Point Validation (CPV), which is a more flexible version of the well-known Central Proximity Score (CPS)~\citep{li2025_screenspotpro} used in prior research.
    \item
    We have publicly released our test dataset to encourage and support future studies in this area.
\end{highlights}

%% Keywords

\end{frontmatter}



%% === SECTIONS
\section{Introduction}\label{intro} 

    Repetitive tasks are activities which require little critical thinking and are regularly repeated in organizations, such as businesses and public administrations. These tasks are often considered tedious and unrewarding by the employees who perform them, which can lead to monotony and professional burnout. To reduce this burden, it becomes important to design an autonomous, AI-powered agent capable of interacting with Graphical User Interfaces (GUIs) to automate tasks. A major challenge in building this agent is enabling it to understand and interact with the GUI environment.

    In this context, many solutions use structured data such as HTML. However, these types of data can be lengthy, impractical, and even inaccessible (e.g., on desktops). To overcome this, it is useful to design an Artificial Intelligence (AI) agent that only relies on screenshots for interactions with its GUI environment.

    Modern GUI visual grounding~\citep{cheng2024seeclick} focuses on aligning natural language instructions with the correct visual targets within desktop GUIs. Given a GUI screenshot $I$ and a natural language instruction $Q$, the objective is to identify the target GUI element $E^*$ that corresponds to the instruction. Each GUI screenshot $I$ contains multiple candidate elements $\mathcal{E} = {E_1, E_2, ..., E_n}$, where each element $E_i$ is defined by its visual region (bounding box) and potentially associated text.

    The recent progress of Large Language Models (LLMs) has opened many possibilities in automating numerous AI tasks. Transformer-based models such as BERT (Bidirectional Encoder Representations from Transformers)~\citep{BERT}, T5~\citep{T5}, and GPT-3 (Generative Pre-trained Transformer)~\citep{GPT3} have achieved great success for Natural Language Processing (NLP) applications, such as text generation~\citep{jiang2023mistral,touvron2023llama}, translation~\citep{moslem2023finetuning, waldendorf2024_translation, wu2024_translation}, and summarization~\citep{BART, zhang2020pegasus, Takeshit2022_X-SCITLDR}. Their success has also inspired new research directions that combine language and vision for more general AI systems.

    More precisely, the use of transformers was later extended to the computer vision domain where researchers revolutionized multiple applications such as image classification~\citep{EfficientNet, BiT, ViT, Beit, CLIP, LeViT, SwinTransformerV2, Perceiver, ConvNet, dinov2}, object detection~\citep{DETR, DeformableDETR, ConditionalDETR, TableTransformer, yolos, DETA} and object understanding~\citep{ZhengyuanYang2019, LinweiYe2019, SeqTR, LAVT} through multi-modal foundation models. An interesting sub-category of object understanding for developing vision capabilities for GUI agents is Referring Expression Grounding (REG) where the key challenge is associating natural language expressions with specific objects or regions in visual data. REG involves understanding and identifying the referred objects or regions based on the contextual information provided by the language, which can include visual attributes, spatial relationships, and interactions with the surrounding environment. This task is pivotal in Natural Language Understanding (NLU) and GUI environment understanding. It bridges the semantic gap between linguistic descriptions and the visual world, enabling machines to understand and interact with the environment more effectively.  

    Research conducted by \cite{Mao2016} demonstrated the efficiency of REG in image captioning tasks. Furthermore, \cite{Anderson2017} illustrated its significance in visual question answering while \cite{LinweiYe2019,LAVT}, and \cite{SeqTR} leveraged cross-modal learning  to segments out an object referred to by a natural language expression from an image. Furthermore, \cite{Anderson2017} illustrated its significance in visual question answering, while \cite{LinweiYe2019, LAVT}, and \cite{SeqTR} leveraged cross-modal learning to segment out an object referred to by a natural language expression from an image. Despite the numerous research works in the context of referring to expression grounding, the majority focuses on natural images. Few works studied the REG in the context of GUI instruction grounding. Indeed, research works like \cite{YangLi_NLPUI_2020, JuliaRozanova_NLPUI2021,zhang2023reinforced, Venkatesh,qian-etal-2024-visual}, and \cite{cheng2024seeclick} propose solutions allowing the localization of screen elements based on natural language instructions. This task is especially useful for building autonomous visual GUI agents designed to automate complex tasks on digital devices. %\\

    In this work, we address natural language instruction grounding through two methods. Our contributions are:
    \begin{itemize}
        \item We propose IVGocr (Instruction Visual Grounding with OCR) dataset \& method: it combines an LLM (ChatGPT~\citep{openai2024gpt4}) with an object detection model (YoloV8~\citep{reis2023realtime}) and an OCR module to enable instruction grounding for GUIs.
        \item We propose IVGdirect (Instruction Visual Grounding Direct) dataset \& method: it is more straightforward, relying on the training of a vision-language multimodal model, based on the work of \cite{yan2023universal}, for the specific task of instruction grounding on GUI screens.
        \item We propose Central Point Validation (CPV) metric, a relaxed variant of the classical Central Proximity Score (CPS) metric \citep{li2025_screenspotpro}. 
        \item Our final test dataset is publicly released to support future research.
    \end{itemize}

\section{Related works}
\label{sec:related_works}

    \textbf{Autonomous GUI Navigation}: Several earlier research works investigated GUI automation tasks for web applications~\citep{pmlr-v70-shi17a, Liu2018, Gur2018} and mobile UI applications~\citep{YangLi_NLPUI_2020, Burns2022, li2023spotlight}. More recently, the significant advancements in the NLP field, propelled by the emergence of Large Language Models (LLMs)~\citep{touvron2023llama, xu2024symbolllm, openai2024gpt4, sun2024corex, wu2024oscopilot}, have shifted the focus toward developing LLM-powered GUI agents. This has become a primary area of interest for many contemporary studies. Indeed, some works proposed to tackle this challenge with a focus on prompt engineering of ChatGPT and ChatGPT-4 for web tasks using in context-learning~\citep{zheng2024synapse} and self-refining~\citep{kim2023language}. Other research works proposed to train LLMs for more specific tasks. In this context, \cite{deng2023mind2web} introduced a two-stage method designed to identify target elements in a given HTML file, while  \cite{gur2024realworld} used programming to allow interactions with websites. Despite these advancements, LLMs are still limited because of their ability to process text only. To overcome this, recent works proposed a vision-based GUI navigation~\citep{NEURIPS2023_Shaw, zhang2024you, hong2023cogagent} using GPT-4V~\citep{yan2023gpt4v, gao2024assistgui}. However, they take metadata as input and do not rely on visual data~\citep{zhang2023appagent, zheng2024gpt4vision}. %\\

    \textbf{Large Vision-Language Models} :  The research interest in processing images and text data simultaneously has been getting more and more attention. In this respect, recent research has seen significant efforts in constructing Large Vision-Language Models (LVLMs) capable of processing both images and text jointly~\citep{liu2023visual, zhu2023minigpt4, ye2023mplugowl, li2023otter}. Cross-modal vision-language tasks have become possible to perform with the integration of vision encoders with LLMs. Indeed, through contrastive learning and masked data modeling on large-scale image-text pairs techniques, models like CLIP~\citep{CLIP}, ALIGN~\citep{Jia2021}, Florence~\citep{Florence}, BEIT3~\citep{BEiT2022}, and Flamingo~\citep{alayrac2022flamingo} demonstrate impressive zero-shot capabilities on vision-language challenges. Other works like DALL-E~\citep{Ramesh2021, ramesh2022hierarchical}, and Stable Diffusion~\citep{Rombach2021} rely on image-caption pairs training for the generation of image content given conditioned on textual prompts.

    More recently, Vision-Language Models (VLMs) have evolved towards generalist systems capable of handling an increasingly broad set of multimodal tasks. This includes the ability to perform fine-grained grounding and reasoning over both text and visual inputs, exemplified in tasks such as Referring Expression Comprehension (REC) \citep{LichengYu2016, ByteTrack, zou2023segment}, Referring Expression Segmentation (RES) \citep{Florence, zhang2023dvis, ByteTrack}, and Referring Video Object Segmentation (R-VOS) \citep{RishiBommasani, touvron2023llama, NingXu}. Furthermore, studies have explored grounding outputs in generated visual content, including providing bounding boxes or region highlights aligned with textual references \citep{wang2023visionllm, bai2023qwenvl, chen2024minigptv, chen2023shikra, peng2023kosmos2}. These advances mark a significant shift towards models that not only interpret multimodal inputs but also generate outputs with fine-grained visual grounding and spatial awareness.

    In line with this progression, the latest generation of Vision-Language Models introduced in 2024 and early 2025 has further expanded the horizons of multimodal understanding. Models such as GPT-4o~\citep{openai2024gpt4ocard} and Gemini 1.5 Pro~\citep{geminiteam2024gemini15unlockingmultimodal} embody the emergence of natively multimodal systems capable of seamlessly processing text, images, and even audio within unified architectures.

    Parallel to these, the open-source ecosystem has witnessed the rise of models such as Qwen-VL-Chat~\citep{bai2023qwenvlversatilevisionlanguagemodel}, MiniGPT-4V~\citep{chen2024minigptv}, LLaVA-1.5~\citep{liu2024_llava1.5}, CogAgent~\citep{hong2023cogagent} and Fuyu-8B~\citep{Adept_Fuyu-8B}, which achieve remarkable improvements in visual reasoning, fine-grained object recognition, and multimodal chain-of-thought capabilities. These models outperform earlier systems on challenging multimodal benchmarks including Measuring Massive Multitask Language Understanding (MMMU) \citep{hendrycks2021measuringmassivemultitasklanguage}, M3Exam \citep{zhang2023m3exammultilingualmultimodalmultilevel}, and MathVista~\citep{lu2024mathvistaevaluatingmathematicalreasoning}, highlighting a clear trend toward more sophisticated, general-purpose vision-language intelligence. As training techniques and datasets continue to evolve, these models are approaching human-level performance on tasks that require the integration of both linguistic and visual understanding.

    Despite these rapid advances, current Vision-Language Models remain predominantly focused on natural image understanding and generation, leaving the specialized domain of Graphical User Interfaces (GUIs) largely unexplored. The unique structure of GUI environments—with their compositional layouts, interactive widgets, and visually repetitive components—requires more fine-tuned grounding capabilities than those needed for natural images. To address this, several recent efforts have been proposed to adapt VLMs for desktop element localization and instruction following. For instance, WinClick \citep{hui2025_winclick2024} and SeeClick \citep{cheng2024seeclick} both focus on grounding natural language instructions in static GUI screenshots, but differ in that WinClick introduces a multitask design using interleaved demonstration-action pairs, whereas SeeClick centers its approach on aligning text to bounding boxes via instruction grounding. Meanwhile, GUI-WORLD \citep{chen2025_guiworld} proposes a large-scale synthetic benchmark for GUI agents, leveraging 3D environments to model real-time interactions, contrasting with DeskVision \citep{xu2025_deskvision}, which builds a multi-turn task-solving benchmark grounded in real-world desktop screenshots and user queries.
    
    Similarly, ScreenSeekeR~\citep{li2025_screenspotpro} takes a retrieval-based approach by formulating instruction grounding as a visual element ranking problem rather than direct object detection. This diverges from other systems like SeeClick or DeskVision, which use end-to-end localization objectives. Despite differences in architectures and objectives, these models share a common goal: enabling instruction-conditioned grounding in digital interfaces—often through adapted versions of VLM backbones, multi-modal alignment heads, and synthetic or semi-synthetic data tailored for GUI environments.

    Despite these rapid advances, current Vision-Language Models remain predominantly focused on natural image understanding and generation, leaving the specialized domain of Graphical User Interfaces (GUIs) largely unexplored. In this paper, we investigate natural language instruction grounding within GUI environments, an emerging and underexplored setting that we believe holds significant potential for the development of autonomous GUI agents capable of perceiving, reasoning, and acting within complex digital interfaces.

\section{Instruction Visual Grounding with OCR (IVGocr)}
\subsection{IVGocr Dataset}
\label{subsec:IVGocr_dataset}

    We propose to build a training dataset composed of annotated GUI screenshots of multiple desktop GUI screens. We devise the following list of the most relevant and frequently occurring targets for object detection training: \say{Button}, \say{Text field}, \say{Text area}, \say{Checkbox}, \say{Radio button}, \say{Text}, \say{Link}, \say{List}, \say{Tab}, \say{Dialog box}, \say{Image}, \say{Progress bar}, \say{Toolbar}, and \say{Menu bar}. The final dataset is composed of $1,264$ images and $102,760$ bounding boxes.
    
    The dataset annotations are produced in a two step fashion. First, an automatic software extracts the metadata of each desktop screen element. This includes the element type, description, and location on screen (bounding box). The second step consists in the manual verification of the correctness of the retrieved information as well as data cleanup. During the data cleanup, only the most relevant GUI screen element types are kept in the dataset. This is done in an effort to remove outliers and assure a smoother training process.

    Note that this dataset was used for training only. The evaluation dataset is taken from the one we propose for our second approach (Subsection~\ref{subsec:IVGdirect_dataset}). The goal is to later enable comparable results across all tested methods.

\subsection{IVGocr Method}

\begin{figure}[!ht]
        \centering
        \begin{tabular}{c}
          \includegraphics[scale=0.35]{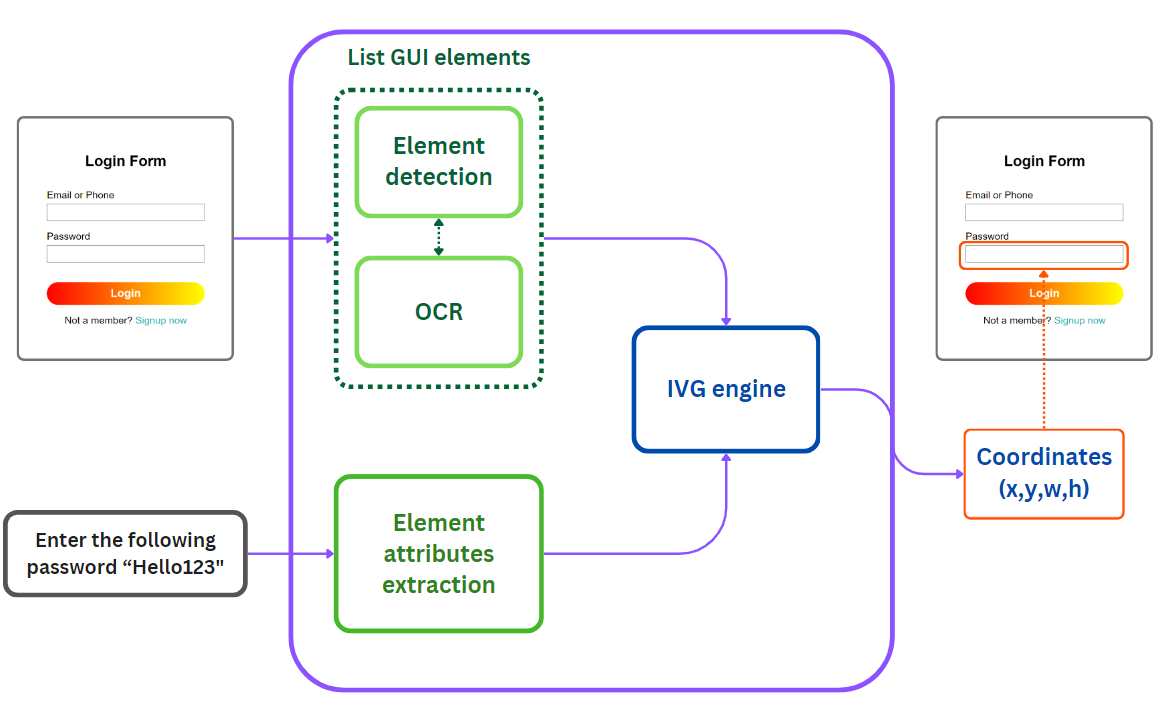} \\  
        \end{tabular}
            \caption{IVGocr architecture. Our 3-step pipeline first detects GUI elements using a YOLOv8 object detector fine-tuned for GUI components. An OCR module reads on-screen text and associates it with detected elements. A large language model (LLM) then extracts the target element type and role from the instruction. Finally, the LLM matches this target to the element list and returns its screen coordinates.}
            \label{fig:SICocr_architecture}
    \end{figure}

    YoloV8 model~\citep{reis2023realtime} demonstrated cutting edge performance in object detection and tracking, instance segmentation, image classification and pose estimation tasks \citep{yolov8_better}. Motivated by its success and the recent advancements in LLMs, we propose a novel 3-step approach that leverages YoloV8's object detection capabilities coupled with an Optical Character Recognition (OCR) module and OpenAI's LLM for GUI instruction grounding able to locate an element on the screen given a natural language instruction from the user and a screenshot of the current GUI screen. Figure~\ref{fig:SICocr_architecture} showcases the proposed IVGocr architecture. Below, we offer a comprehensive breakdown of each stage within the IVGocr instruction grounding methodology.

\subsubsection{GUI Element Detection and Text Recognition}
    \label{list_comp_gui}
    
    The first step of the approach consists in listing all the elements present in the GUI (i.e. tab, button, text field, list, etc.). To this end, we propose to deploy YoloV8~\citep{reis2023realtime} for object detection and fine-tune it for GUI objects. 
    
    In spite of the good accuracy of the object detection module, it still exhibits some limitations due to its inability to read text and correctly identify each element in an image. For example, it would not be able to distinguish between the \say{Cancel} button and the \say{Submit} button and would simply identify them as \say{Button}. For this reason, we propose adding an OCR module to help identify each element on the GUI screen. To this end, we design the OCR module to first \say{read} all the text present on the screen and return the coordinates of each text. We then match the coordinates of each object on the screen with the corresponding text. This process allows us to identify for each element its type (ex.: botton) and role (ex.: cancel) and it finally returns a list containing the Id, type, role, and coordinates of each element in the GUI.
    
\subsubsection{Instruction Parsing for Target Element Identification}
\label{name_type_extract}
    
    The second step of the architecture is dedicated to extract information about the target element on the screen from the input user instruction written in natural language. More specifically, we aim to know the type and role of the element we are looking for on the screen. For example, given the following instruction: \say{Please type John in the name field}, we would have a type corresponding to \say{text field} and a role corresponding to \say{name}. To this end, we resort to prompt engineering using OpenAI's GPT-3.5 Turbo LLM~\citep{openai2024gpt4}.

\subsubsection{Element Matching and Grounding}

    This stage relies on our IVG engine. As described in Sections~\ref{list_comp_gui} and~\ref{name_type_extract}, the first two steps produce, respectively, the list of GUI elements and the information about the target element. Based on these results, the third and final step is designed to return the coordinates of the target element. To achieve this, we leverage OpenAI's GPT-3.5 Turbo LLM through prompt engineering to find the element description in the list that best matches the target. A specific prompt, detailed in the Table below, is used to guide the LLM in the matching process. This prompt has been refined through extensive experimentation and was found to produce the most reliable results. Once identified, the LLM returns the selected element from the list, including its ID, type, role, and coordinates on the GUI screen.

\begin{tcolorbox}[colback=gray!5, colframe=black!70, title=LLM      Prompt: GUI Element Selection, sharp corners, boxrule=0.5pt]
An HMI (human-machine interface) graphical interface contains multiple components on the screen. The following list contains information about these components. Each element in the list is composed of the \texttt{Id}, coordinates \texttt{[x\_min, y\_min, x\_max, y\_max]} of the component in the HMI, its \texttt{type} (button, link, text field, text, ...), its \texttt{value}, and also the info of the nearest text component (\texttt{top}, \texttt{bottom}, \texttt{left}, \texttt{right}):
\vspace{0.5em}

\texttt{components list = <elements\_list>}
\vspace{0.5em}

Please choose the \texttt{Id} of the element in the components list where this action would be executed: \texttt{<element>}
\vspace{0.5em}

Use each component's info, as well as the components to its right, left, top, and bottom on the HMI to find the best match.
\vspace{0.5em}

The response should follow this format:
\vspace{0.5em}

\texttt{Id : <Id>}
\end{tcolorbox}

 Following this 3-step strategy, we are able to achieve instruction grounding for the GUI environment. Despite the good performance of this approach, it shows some limitations. For instance, it heavily relies on text information for element identification and would not be able to accurately identify image elements and icons. Furthermore, it calls for the use of OpenAI's GPT-3.5 Turbo LLM which can present a few issues. Namely, data security concerns since it is not an LLM that can be deployed locally, as well as potential financial concerns. To address these issues, we propose a more local and straightforward approach in what follows.

\section{Instruction Visual Grounding direct (IVGdirect)}
\subsection{IVGdirect Dataset}
  \label{subsec:IVGdirect_dataset}
    %Regarding the training dataset, 
    Due to the lack of training datasets tailored for visual grounding for GUI screens and more specifically desktop GUI screens, we propose to create an extensive desktop dataset containing image-expressions pairs. The dataset is designed so that each element of the screen is referenced by a variety of expressions that the user might use to locate the element. Each image-expression pair corresponds to a visual element and a list of expressions associated with it. For example, a submit button could be paired with the following expressions list: [\say{submit button}, \say{submit}, \say{button to submit}]. We also divide the dataset into training, validation, and testing sets using an 80-15-5 split, respectively. We publicly release our test set for future experiments \footnote{\url{https://huggingface.co/datasets/Novylab/GUI_grounding}}.

\begin{figure}[!ht]
        \centering
        \begin{tabular}{c}
          \includegraphics[scale=0.3]{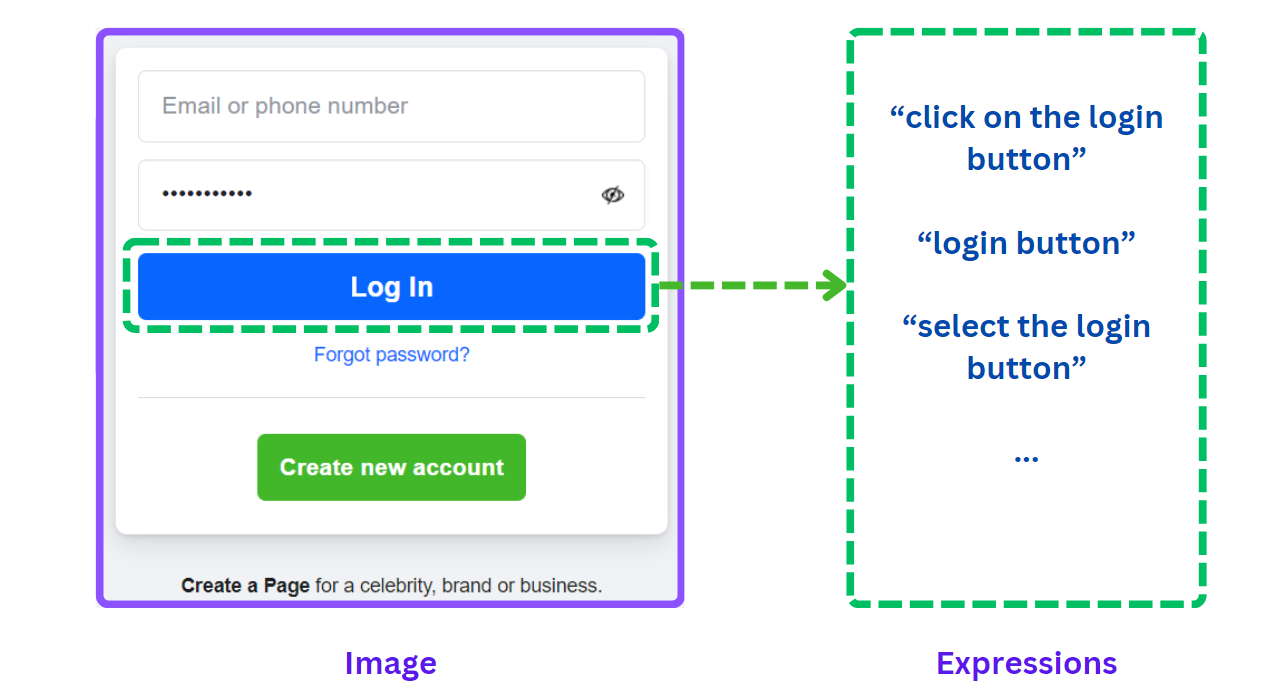} \\
        \end{tabular}
        \caption{Example of an image-expressions pair from IVGdirect Dataset.}
        \label{fig:data_example}
    \end{figure}

    The GUI screen element information is collected with the same process used for the IVGocr approach. It is worth pointing out that the used dataset for the IVGdirect approach is relatively large. We opted for a smaller test split to optimize time for the experiments and more specifically for the comparison with the state of the art. Figure~\ref{fig:data_example} presents an example of an image-expressions pair of the dataset.
    
    Our training set contains $51,433$ image-expression pairs. We categorize the elements into $7$ categories: Button, Tab, Link, Text field, Checkbox, Radio button, and List. We aimed to achieve a balanced distribution as much as possible, but due to a lack of certain elements, the dataset remains imperfectly balanced.
    
    Our test set is composed of $2,464$ image-expression pairs and each pair can be divided into 8 categories: `Button', `Text field', `Text area', `Checkbox', `Radio button', `Link', `List', and `Tab'. Note that the category `Text area' only appears in the test set because of lack of examples. Table~\ref{tab:merged_data_dist} shows the category distribution of the dataset.

    \begin{table}[!ht]
        \centering
        \small
        \begin{tabular}{l c c | c c}
            \hline
            \multicolumn{3}{c|}{Train stats} & \multicolumn{2}{c}{Test stats} \\
            \hline
            Category & Number of samples & Percentage \% & Number of examples & Percentage \%  \\
            \hline\hline
            Button            & 9156  & 17.80\%  & 519  & 21.06\%  \\
            Tab               & 10361 & 20.14\%  & 574  & 23.30\%  \\
            Text field/area   & 9480  & 18.43\%  & 439  & 17.82\%  \\
            Link              & 8435  & 16.40\%  & 367  & 14.89\%  \\
            Checkbox          & 7994  & 15.54\%  & 320  & 7.75\%   \\
            Radio button      & 4094  & 7.96\%   & 191  & 12.99\%  \\
            List              & 1913  & 3.72\%   & 46   & 1.87\%   \\
            Text area         & -     & -        & 8    & 0.32\%   \\
            \hline
        \end{tabular}
        \caption{IVGdirect Dataset distribution across categories.}
        \label{tab:merged_data_dist}
    \end{table}

\subsection{IVGdirect Method}

    \subsubsection{Motivation}

    The GUI grounding task is particularly challenging due to the diversity of visual styles, the presence of non-standardized icons, and the inherent ambiguity in natural language referring expressions. Addressing these challenges requires robust multimodal models capable of understanding both visual and linguistic cues.

    Inspired by the work of \cite{yan2023universal}, which introduces a vision-language multimodal instance perception model, we propose to address the challenge of grounding the desktop GUI screen. In fact, in the context of natural images, \cite{yan2023universal} demonstrate good performance for different types of perception tasks, including but not limited to object detection, instance segmentation, and visual grounding of natural language. However, transferring the visual grounding capabilities of \cite{yan2023universal} to the GUI screen element grounding context proves to be a challenging task that very few recent works have addressed. This is in part due to the vastly different distribution of the synthetic images corresponding to the GUI screen compared to natural images. Furthermore, elements on a GUI screen present different properties from natural objects in a natural image. For instance, elements on a GUI screen tend to have more rigid horizontal and vertical borders as opposed to the more dynamic diagonal borders present in natural images. Additionally, while natural objects tend to be rich in texture, GUI elements can be more ``flat'' or homogeneous, which is a major characteristic specific to synthetic images. For this reason, the visual features of a synthetic GUI screen image are distinctly different from the natural image features. In this respect, a specialized multimodal vision-language model is necessary to treat the unique case of instruction visual grounding for GUI screens.

    \begin{figure}[!ht]
        \centering
        \begin{tabular}{c}
          \includegraphics[scale=0.35]{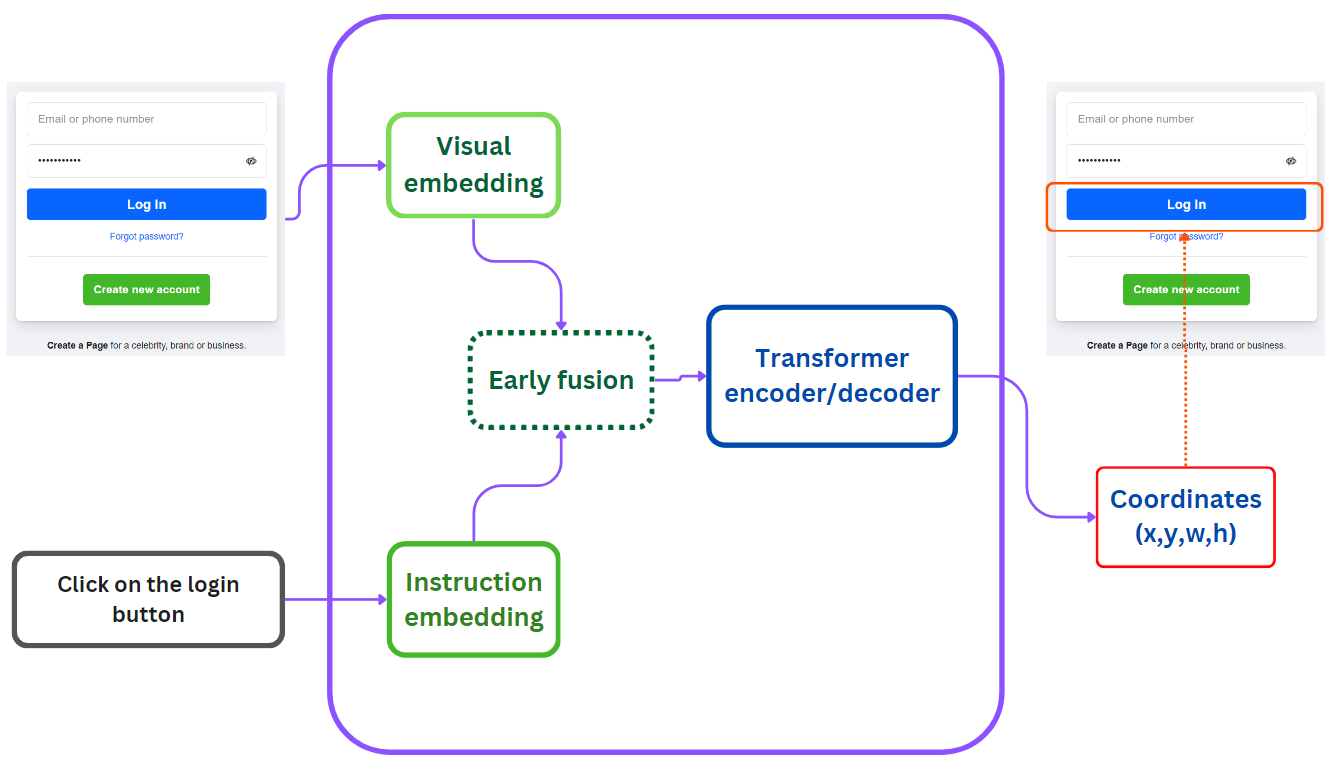} \\ 
        \end{tabular}
            \caption{IVGdirect architecture. The model uses a BERT encoder for instruction embedding and a ResNet-50 backbone for visual feature extraction. A fusion module integrates both modalities, and a  Deformable DETR transformer retrieves the most relevant GUI element guided by the instruction.}
        \label{fig:SICdirect}
    \end{figure}
    
    In the following, we provide a detailed description of the multimodal model architecture we use for GUI visual grounding.

    \subsubsection{Overview of the Architecture}
    \label{subsec:Global_approach_description}

    Usually, an LVLM-based model $\mathcal{M}$ for GUI visual grounding consists of a visual encoder $F_v$ and a language encoder $F_t$ that project the image $I$ and instruction $Q$ into a shared embedding space. A matching module $C$ computes the relevance scores between the instruction and each candidate element. The output is a predicted target element $\hat{E} = \arg\max_{E_i \in \mathcal{E}} p(E_i|I,Q)$, where $p(E_i|I,Q)$ denotes the grounding score for element $E_i$. We get inspiration from the current problem definition and formulate it as a natural-language-instruction-guided object discovery and retrieval problem. The model calls for a 4-step architecture: instruction embedding generation, visual embedding generation, image-instruction feature fusion, and object discovery and retrieval. Figure~\ref{fig:SICdirect} shows the global architecture of the approach.

    \paragraph{Instruction embedding generation} 
    The first step consists in the generation of the natural language instruction embedding. To this end, we use the BERT language encoder introduced in \cite{BERT}. This embedding is consequential for guiding the object grounding process.

    \paragraph{Visual embedding generation} Concurrently with the language embedding generation, the current GUI screen undergoes processing through a visual encoder to obtain hierarchical visual features. The visual encoder uses ResNet-50 introduced in \cite{ResNet50} as the backbone.
    
    \paragraph{Image-language Feature Fusion}
    An early fusion module then enhances the visual features with the natural language instruction embedding to enable deep information exchange and produce highly discriminative representations for instance prediction.
    
    \paragraph{Object discovery and retrieval} 
    Utilizing enhanced visual and language representations, this phase employs a Transformer-based architecture  \citep{attention_is_all_you_need} to generate instance proposals, which are then filtered based on language-instance matching scores to retrieve the final object instances. From this perspective, we adopt the encoder-decoder architecture proposed by Deformable DETR \citep{DeformableDETR} due to its flexible and efficient architecture. The Transformer encoder utilizes hierarchical, instruction-aware visual features as input. Leveraging the efficient Multi-scale Deformable Self-Attention mechanism \citep{DeformableDETR}, it enables comprehensive exchange of target information across various scales, resulting in enhanced instance features for the subsequent decoding phase.  
    This architecture allows for a relatively simple and direct approach for desktop GUI element grounding.

    \section{Experimental settings}

    \subsection{Compared Datasets}
    In addition to our IVGdirect dataset, we use the following datasets for comparison:

    \textit{\textbf{ScreenSpot}}~\citep{cheng2024seeclick} - is a benchmark designed specifically for GUI environments. It comprises over 1,200 natural language instructions collected from diverse platforms including iOS, Android, macOS, Windows, and web-based environments, with each instance annotated with element types such as text or icon/widget components, all in English. For the desktop subset, the benchmark focuses on screenshots from Windows and macOS systems. Annotators were instructed to select commonly used applications and everyday operations to ensure the dataset reflects realistic and practical use cases.

    To build the dataset, annotators first captured screenshots based on typical daily device usage. Once the screenshots were collected, they annotated key interactive elements—primarily clickable regions—relevant to user interactions. Finally, they composed natural language instructions aimed at prompting models to interact with specific annotated elements. In our experiments, we specifically focused on the desktop subset of ScreenSpot, leveraging only the data from Windows and macOS platforms for instruction grounding and evaluation.

    \textit{\textbf{ScreenSpot-Pro}} ~\citep{li2025_screenspotpro} -  is an extension of the original ScreenSpot benchmark, specifically designed to capture high-resolution screenshots from professional desktop environments. Unlike the original dataset, which covers multiple devices and general applications, ScreenSpot-Pro focuses exclusively on desktop platforms and includes six professional software categories: Development and Programming, Creative Software, CAD and Engineering, Scientific and Analytical tools, Office Software, and Operating System Commons. This broader coverage introduces more complex visual layouts and realistic scenarios for evaluating vision-language models in GUI grounding tasks.

    In our work, we specifically use the Operating System Commons (OS) subset of ScreenSpot-Pro, which focuses on general-purpose system environments such as file management, desktop navigation, and utilities. This choice aligns with our research, which targets the accurate localization of elements within desktop GUI environments—an essential step for enabling instruction grounding and interaction in real-world computing contexts.

\subsection{Baseline methods}
We compare our two methods with the following ones:
    \begin{itemize} 
        \item 
        \textbf{\textit{SeeClick}} -  a scalable vision-language pre-training framework \citep{cheng2024seeclick}—uses a modular transformer trained on large-scale image-text pairs without object-level annotations. Pre-training combines masked language modeling with region-word alignment. Fine-tuning employs all referring expression datasets with cross-modal grounding objectives. A filtering mechanism selects informative samples, and hyperparameters are tuned once and kept fixed for downstream tasks.

        \item 
        \textbf{\textit{CogAgent}} - a vision-language-action foundation model \citep{hong2023cogagent}—employs dual-stream transformers with cross-modal fusion, trained on diverse multimodal datasets spanning vision-language understanding and embodied reasoning. Knowledge distillation preserves past capabilities, while task prompts enable multi-task inference without further fine-tuning. Hyperparameters remain fixed throughout evaluation.

        \item 
        \textbf{\textit{Fuyu}} - a family of open multimodal transformers \citep{Adept_Fuyu-8B} — uses a unified backbone for vision and language inputs, pre-trained with contrastive learning and fine-tuned on instruction-following datasets. Zero-shot generalization is achieved via tokenized multi-modal inputs, with selective replay maintaining prior knowledge. Training hyperparameters are frozen post pre-training for consistent evaluation.

        \item 
        \textbf{\textit{GPT-4o}} - a natively multimodal model introduced by OpenAI \citep{openai2024gpt4ocard}—unifies text, vision, and audio inputs within a single, end-to-end transformer architecture. Unlike previous GPT-4 variants relying on separate encoders, GPT-4o jointly processes modalities without adapters or late fusion. The model is trained with a mixture of supervised fine-tuning and reinforcement learning from human feedback (RLHF), covering a wide range of natural and multimodal tasks. GPT-4o achieves strong performance across vision language benchmarks, including fine-grained localization and reasoning, while offering low latency and enhanced efficiency for real-time interaction.

    \end{itemize}

\subsection{Evaluation Metrics}

\textbf{- Intersection over Union (IoU)}~\citep{IoU} - it measures the overlap between the predicted bounding box and the ground truth bounding box. The IoU score ranges from 0 to 1, where 0 means there is no overlap (i.e., false prediction) and 1 indicates perfect overlap (i.e., perfect prediction). Formally, the IoU between a predicted bounding box $B_p$ and a ground truth bounding box $B_{gt}$ is defined as:

    \begin{equation}
    IoU(B_p, B_{gt}) = \frac{|B_p \cap B_{gt}|}{|B_p \cup B_{gt}|}
    \label{Equation 1}
    \end{equation}

\vspace{1em}

\hspace{-2em}
\textbf{- Central Proximity Score (CPS)} \cite{li2025_screenspotpro} - In the case of GUI agents, the center of the predicted bounding box is particularly important. It helps determine whether the predicted location is within the correct target area. However, the IoU metric only measures the overlap between two boxes and does not consider the position of the center point. For example, a prediction with a low IoU (e.g., 0.4) might still have its center in the correct region. To better evaluate such cases, we use the CPS metric. The latter assigns a higher score to predictions whose centers lie closer to the center of the ground truth box, using a normalized Gaussian function (Equation \ref{eq:score_function}).

\begin{equation}
    CPS(B_p, B_{gt}) = S(x, y) =
    \begin{cases}
        \exp\left( -\dfrac{(x' - 0.5)^2 + (y' - 0.5)^2}{2\sigma^2} \right), & \text{if point inside} \\
        0, & \text{otherwise}
    \end{cases}
    \label{eq:score_function}
\end{equation}

where:
\begin{equation}
    x' = \dfrac{x - x_1}{x_2 - x_1}, \quad
    y' = \dfrac{y - y_1}{y_2 - y_1}
    \label{eq:normalized_coords}
\end{equation}

\vspace{1em}

where $(x, y)$ is the center of the predicted box, and $(x_1, y_1, x_2, y_2)$ represent the coordinates of the ground truth box. As in \cite{li2025_screenspotpro}, we set $\sigma$ to 0.3.

This metric is particularly relevant for GUI task automation, where actions such as clicks or text inputs are typically executed at the center of the target element. Therefore, ensuring that the predicted center falls within the correct region is essential for reliable interaction.

\vspace{1em}

\hspace{-2em}
\textbf{- Central Point Validation (CPV)} - We propose a new metric called Central Point Validation (CPV) that checks whether the center of the predicted bounding box is inside the ground truth box. It can be seen as a relaxed version of CPS metric, and it is defined in Equation \ref{cpv_metric}:
    
    \begin{equation}
        CPV(B_p, B_{gt}) =
        \begin{cases}
        1, & \text{if } \text{center}(B_p) \in B_{gt} \\
        0, & \text{otherwise}
        \end{cases}
        \label{cpv_metric}
    \end{equation}
    
We report the percentage of correct predictions across the test dataset.

\subsection{Training Protocol}

\begin{itemize}
    \item 
    \textbf{\textit{IVGocr Method}} - To optimize object detection performance, the fine-tuning process of YoloV8 is meticulously undertaken, with a particular focus on hyper-parameter settings, including the batch size, learning rate and weight decay. By opting for a batch size of $16$, we aim to enhance both computational efficiency and the model's learning capabilities. After empirical experimentation with different learning rate values, we conclude that a learning rate of $0.01$ is best suited for an optimal training in terms of speed as well as performance. Furthermore, in an effort to mitigate overfitting, we use weight decay as it reduces model complexity and variance. In this regard, we use a small weight decay of $0.0005$ as it allows for improvement in validation. This careful selection of hyper-parameters is found to not only speed up the learning process but also ensure ongoing refinement throughout the training phase and facilitate effective model optimization. We conduct the training on a A6000 GPU equipped with a $48$ GB of GDDR6 memory for a total of $400$ epochs. %\\
    
    \item 
    \textbf{\textit{IVGdirect Method}} - We conduct the training of the model with special care and attention to the configuration of the hyper-parameters, namely the batch size and learning rate to achieve optimal performance for expression grounding. Indeed, we choose a batch size of $32$ to maximize the computational and learning efficiency. It should be noted that we aimed to maximize the batch size given the computational capabilities we have at our disposal. Moreover, we select a learning rate of $2e-4$. This decision comes after extensive empirical inspection of the effect of the learning rate on the training process. This value is found to be the best suited to ensure model optimization while boosting convergence speed. This hyper-parameter configuration can guarantee not only rapid learning but also continuous refinement during training.
\end{itemize}

\section{Experiments and Results}

    In this section, we present experimental results for our proposed approaches, IVGocr and IVGdirect, evaluated on the IVGdirect dataset and two state-of-the-art benchmarks: ScreenSpot and ScreenSpot-Pro. We use IoU and CPV to assess performance on our dataset, and the CPS metric for fair comparison on the public benchmarks. The results highlight the advantages of domain-specific grounding methods over general-purpose vision-language models in GUI understanding tasks.

    \subsection{IVGdirect dataset}
    
    Table~\ref{tab:eval_category} depicts the category-wise evaluation results of the state-of-the-art methods GPT-4o~\citep{openai2024gpt4ocard}, Fuyu~\citep{Adept_Fuyu-8B}, CogAgent~\citep{hong2023cogagent}, SeeClick~\citep{cheng2024seeclick}, as well as the proposed IVGocr and IVGdirect approaches. As we can see, CogAgent and the IVGdirect method consistently outperform the rest across most GUI element categories, both in terms of Intersection over Union (IoU) and Central Point Validation (CPV), with IVGdirect achieving the highest global performance in CPV (79.24\%) and nearly matching CogAgent in IoU (65\% vs. 67\%).

    \begin{table}[!ht]
        \centering
        \small
        \setlength{\tabcolsep}{0.5em}
        \begin{tabular}{c c c c c c c}
            \hline
            \multicolumn{7}{c}{\textbf{IoU}} \\
            \hline
            Category    &  GPT-4o & Fuyu & CogAgent & SeeClick & IVGocr & IVGdirect \\
            \hline
            Tab          & 0.36 & 0.35  & \textbf{0.95}  & - & 0.40  & 0.93 \\
            Button       & 0.15 & 0.11  & \textbf{0.72} & - & 0.16  & 0.67 \\
            Text field   & 0.21 & 0.19  & 0.70 & - & 0.24  & \textbf{0.74} \\
            Link         & 0.21 & 0.15  & \textbf{0.81}  & - & 0.15  & 0.59 \\
            Radio button & 0.12 & 0.01  & 0.25  & - & 0.06  & \textbf{0.31} \\
            Checkbox     & 0.01 & 0.02  & 0.31 & - & 0.04  & 0.37 \\
            List         & 0.19 & 0.26  &  \text{0.77} & - & 0.37  & 0.75 \\
            Text area    & 0.49 & 0.51  & \textbf{0.85}  & - & 0.67  & 0.84 \\
            \hline
            Global       & 0.21 & 0.20 &  \textbf{0.67} & - & 0.26   & 0.65 \\
            \hline
            \\
            \hline
            \multicolumn{7}{c}{\textbf{CPV}} \\
            \hline
            Category       & GPT-4o & Fuyu & CogAgent & SeeClick   & IVGocr & IVGdirect \\
            \hline
            Tab            &  29.51\% &  31.24\% & \textbf{98.20\%} & 32.41\%   & 58.36\% & 97.04\% \\
            Button         &  24.23\% &  24.25\% & \textbf{90.31\%} & 26.01\%  & 22.74\% & 85.55\% \\
            Text field     &  28.26\% &  31.42\% & 87.21\% & 14.35\%  & 33.94\% & \textbf{91.34\%} \\
            Link           &  42.32\% &  49.25\% & \textbf{75.22\%}& 44.69\%  & 47.68\% & 71.12\% \\
            Radio button   &  8.25\% &  11.25\% & 37.27\% & 7.85\%   & 10.99\% & \textbf{41.36\%} \\
            Checkbox       &  1.01\% &  1.25\% & \textbf{55.20\%} & 1.56\%   & 3.12\%  & 49.69\% \\
            List           &  10.86\% &  12.36\% & 87.55\% & 15.22\%  & 45.65\% & \textbf{97.83\%} \\
            Text area      &  11.41\% &  11.25\% & \textbf{100\%} & 12.51\%   & 87.52\%  & \textbf{100\%} \\
            \hline
            Global         & 19.48\% &  21.53 &  78.87 & 19.32\% & 38.75\% & \textbf{79.24\%} \\
            \hline
        \end{tabular}
        \caption{Category-wise evaluation results on IVGdirect dataset using IoU and CPV metrics}
        \label{tab:eval_category}
    \end{table}
    
    The superior performance of CogAgent, especially in categories such as Tab, Button, Link, and Text area, can largely be attributed to its substantial model size and advanced vision-language alignment capabilities. With billions of parameters, CogAgent leverages deep multimodal understanding, allowing it to interpret complex spatial and textual cues present in GUI screenshots. This makes it particularly adept at semantically grounding queries to their corresponding UI elements, achieving up to 98.2\% CPV on Tabs and 100\% on Text Areas.

    In contrast, GPT-4o and Fuyu, although powerful in general-purpose vision-language tasks, fall short in GUI grounding tasks, especially on categories like Checkbox and Radio Button—elements that are visually minimal and structurally ambiguous. These models generally prioritize textual reasoning and broad image understanding, which may not translate effectively to fine-grained, structured visual contexts such as those in user interfaces.

    The IVGdirect method, which achieves the best global CPV among all models at 79.24\% and a close second-best global IoU of 65\%, demonstrates the effectiveness of direct visual grounding pipelines tailored specifically for GUI tasks. Unlike large generalist models, IVGdirect is designed with domain-specific strategies, such as the incorporation of GUI-specific features. This targeted approach gives it an edge, especially in more structured elements like Text field and List, where it even surpasses CogAgent in CPV (91.34\% vs. 87.21\% on Text field and 97.83\% vs. 87.55\% on List).
    
    Another relevant observation is the performance gap between IVGocr and IVGdirect. While both are proposed approaches, IVGocr lags behind in both metrics. This difference suggests that OCR-based pipelines, which rely heavily on text extraction followed by semantic matching, may not be sufficient for precise element localization—especially when visual layout and non-textual cues (like icons or spacing) are critical. IVGdirect’s strong results on Checkboxes and Radio Buttons, categories where textual content is sparse or absent, further support this hypothesis.

    As for SeeClick, the method demonstrates modest performance in CPV, but lacks IoU results due to its output format: SeeClick returns a single point of interest rather than a bounding box. This inherently limits its spatial accuracy and complicates direct comparisons with other methods. Moreover, Figures~\ref{fig:seeclick_age_field} and \ref{fig:seeclick, Text field : nom} in Appendix~\ref{app:example_wrong_pred} highlight common failure modes of SeeClick. These examples show that the model often fixates on text surrounding the actual GUI element, which causes it to miss the target area entirely. While leveraging contextual text can be a smart strategy, relying on a single coordinate point rather than a bounding box reduces the chance of accurately capturing the full element, especially in dense or visually complex interfaces.
    
    In summary, the results underscore the value of domain-specific model design and dataset alignment. CogAgent's large-scale architecture provides impressive performance, but the IVGdirect approach achieves comparable—and often better—results using a more targeted methodology, showcasing the potential of focused, efficient models for GUI grounding tasks.

    \subsection{ScreenSpot and ScreenSpot-Pro Datasets}

    To fairly evaluate our proposed approaches against the state-of-the-art, we conducted experiments on two established benchmarks: ScreenSpot and ScreenSpot-Pro (OS), using the Central Proximity Score (CPS) as the evaluation metric. Table~\ref{tab:results_combined} presents the results for leading vision-language models (LVLMs)—GPT-4o, Fuyu, CogAgent, SeeClick, IVGocr, and IVGdirect—on these datasets. Designed to test grounding capabilities in screen-based scenarios, both benchmarks involve localizing text, icons, and widgets based on natural language descriptions. The results reveal substantial performance disparities, offering key insights into model generalization and robustness under increasingly complex conditions.
    
    \begin{table}[!h]
        \centering
        \small
        \begin{tabular}{lccccccc}
            \toprule
            \multirow{2}{*}{\textbf{LVLMs}} &
            \multicolumn{3}{c}{\textbf{ScreenSpot}} &
            \multicolumn{3}{c}{\textbf{ScreenSpot-Pro (OS)}}  \\
            \cmidrule(lr){2-4} \cmidrule(lr){5-7}
            & \textbf{Text} & \textbf{Icon/Widget} & \textbf{Avg} 
            & \textbf{Text} & \textbf{Icon/Widget} & \textbf{Avg} 
            & \\
            \midrule
            GPT-4o     & 30.1\% & 21.5\% & 25.8\%  & 0.0\%  & 0.0\% & 0.0\%  \\
            Fuyu       & 33.9\% & 4.4\%  & 19.1\%  & 1.8\%  & 0.0\% & 0.9\%  \\
            CogAgent   & \textbf{70.4\%} & 28.6\% & \textbf{49.5\%}  & \textbf{5.6\%}  & 0.0\% & \textbf{2.8\%}  \\
            SeeClick   & 55.7\% & 32.5\% & 44.1\%  & 2.8\%  & 0.0\% & 1.4\%  \\
            \textit{IVGocr}   & 51.2\%  & 30.2\% & 40.7\% &  1.8\%  & 0.0\% & 0.9\%  \\
            \textit{IVGdirect}& 59.3\%  & \textbf{35.1\%}  & 47.2\% & 3.8\%  & 0.0\%  & 1.9\%    \\
            \bottomrule
        \end{tabular}
        \caption{Comparison of LVLMs on the ScreenSpot and ScreenSpot-Pro (OS) datasets using CPS metric.}
        \label{tab:results_combined}
    \end{table}

    On the ScreenSpot dataset, which features a balanced mix of textual and graphical interface elements, CogAgent emerges as the top performer with 70.4\% accuracy on text elements and 28.6\% on icon/widget localization, yielding an overall average of 49.5\%. This demonstrates the strength of its multimodal reasoning, likely enabled by its high-capacity architecture and extensive pretraining. IVGdirect follows closely with 59.3\% on text and a leading 35.1\% on icons/widgets (average: 47.2\%), showcasing the effectiveness of layout-aware grounding strategies tailored for GUI elements. The gap between text and icon performance across models reaffirms that even top-performing LVLMs still struggle with purely visual elements lacking textual anchors.

    SeeClick achieves 55.7\% on text and 32.5\% on icons/widgets (average: 44.1\%), slightly behind CogAgent and IVGdirect. Despite its point-based design and relatively modest architecture, it continues to perform well on visually sparse targets—although its lack of bounding-box precision may limit utility in fine-grained GUI understanding.
    
    Among the remaining methods, IVGocr scores 51.2\% on text and 30.2\% on icons/widgets (average: 40.7\%), placing it behind the top three. As expected, OCR-based approaches like IVGocr handle text-rich UIs well but remain less effective for non-textual elements.
    
    General-purpose LVLMs lag significantly. GPT-4o achieves 30.1\% on text and 21.5\% on icons/widgets (average: 25.8\%), while Fuyu performs worse with 33.9\% on text and only 4.4\% on icons/widgets (average: 19.1\%). These results highlight the limitations of generalist models pretrained on natural images and web captions when applied to structured digital interfaces.
    
    On the more challenging ScreenSpot-Pro (OS) dataset—which includes less text, smaller or denser icons, and more visually cluttered OS-level interfaces—performance across all models drops dramatically. CogAgent remains the best performer with 5.6\% on text and 0.0\% on widgets (average: 2.8\%), followed by IVGdirect at 3.8\%/0.0\% (avg: 1.9\%) and SeeClick at 2.8\%/0.0\% (avg: 1.4\%). IVGocr and Fuyu each score 1.8\% on text and 0.0\% on widgets (avg: 0.9\%), while GPT-4o performs the worst overall with 0.0\% across all categories.
    
    This steep decline highlights a critical limitation: lack of robustness to domain shift. The transition from structured app/web interfaces in ScreenSpot to the heterogeneous, real-world content of ScreenSpot-Pro introduces system-level elements such as dialogs, toolbars, and embedded media that current LVLMs are ill-equipped to interpret. Even models designed for GUI grounding fail to generalize effectively.
    
    In summary, while models like CogAgent and IVGdirect demonstrate strong results on semi-structured screen data, their poor performance on ScreenSpot-Pro underscores a serious gap in current LVLM capabilities. These findings reinforce the need for robust, layout-aware grounding models trained on diverse, real-world screen content to ensure reliable performance across device types, applications, and operating systems.

\section{Conclusion}

    This work tackles the problem of grounding natural language instructions within GUI screenshots—a crucial step toward enabling intelligent, screen-aware agents. To this end, we proposed two methods: IVGocr, which combines object detection with OCR-driven semantic alignment, and IVGdirect, a layout-aware visual grounding approach that leverages universal instance perception models to directly localize relevant interface elements.

    Through extensive evaluation on three diverse benchmarks—IVGdirect, ScreenSpot, and ScreenSpot-Pro—we demonstrate that IVGdirect consistently outperforms or rivals state-of-the-art vision-language models, including GPT-4o, Fuyu, and SeeClick. It achieves the highest global CPV (79.24\%) on the IVGdirect dataset and performs competitively with the much larger CogAgent, surpassing it on several structured categories like text fields and lists. These results underscore the benefits of domain-specific architectural design for GUI understanding, especially in contrast to generalist LVLMs trained on natural images and web data.

    Our findings also reveal key limitations of OCR-based models such as IVGocr, which struggle with visually minimal or non-textual UI elements, as well as models like SeeClick, whose point-based output restricts spatial precision. Moreover, on the more realistic and challenging ScreenSpot-Pro benchmark—characterized by OS-level screens with visual clutter and fewer textual cues—all models, including IVGdirect and CogAgent, exhibit a steep drop in performance. This clearly exposes the lack of robustness to domain shift and highlights a broader gap in current screen grounding capabilities.

    To address these limitations, future research should focus on enhancing generalization across heterogeneous interface types and real-world screen content. Key directions include: domain adaptation strategies to mitigate performance degradation under distribution shifts, integration of semantic scene graphs or hierarchical layout models to improve understanding of compositional GUI structure, and interactive feedback mechanisms that allow agents to refine predictions based on user input or downstream execution results.

    \paragraph{\textbf{Acknowledgements}} 

    We would like to acknowledge Novelis for their support in publishing this article. We are especially grateful for the assistance and contributions of their research team.

%% ======= APPENDIX
\clearpage
\appendix
\section{Examples of wrong predictions with SeeClick}
\label{app:example_wrong_pred}

    \begin{figure}[!ht]
            \centering
            \begin{tabular}{c}
              \includegraphics[scale=0.3]{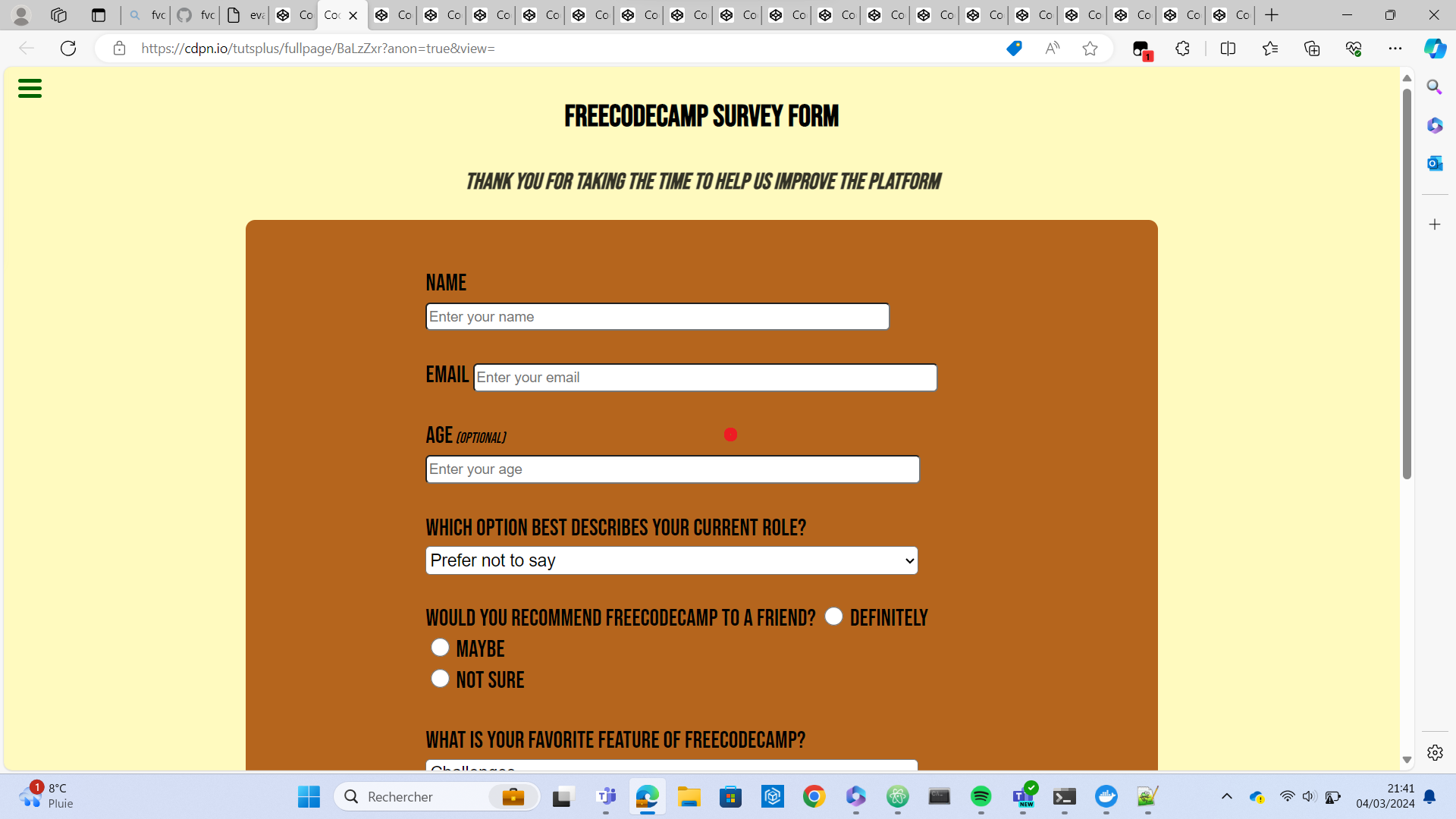}  
              \\ 
            \end{tabular}
            \caption{Approach: SeeClick. Reference text: \textquote{Age field}.}
            \label{fig:seeclick_age_field}
        \end{figure}
    
    \begin{figure}[!ht]
        \centering
        \begin{tabular}{c}
          \includegraphics[scale=0.3]{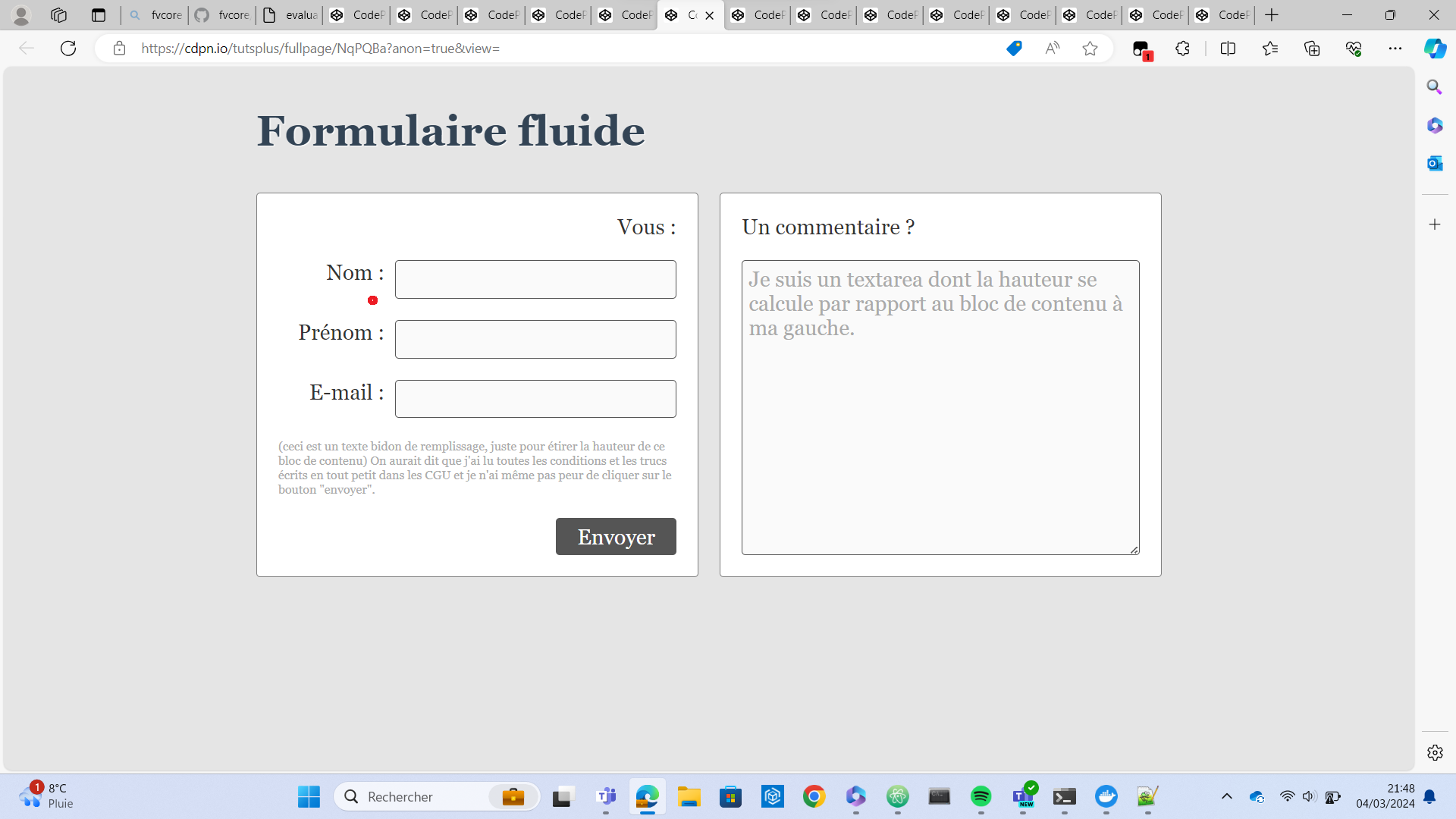}\\
        \end{tabular}
        \caption{Approach: SeeClick. Reference text: \textquote{Text field : nom}.}
        \label{fig:seeclick, Text field : nom}
    \end{figure}
    \clearpage
    
    \section{Examples of wrong predictions with IVGdirect.}
    \label{app:examples_wrong_IVGdirect}
    
    \begin{figure}[!ht]
            \centering
            \begin{tabular}{c}
              \includegraphics[scale=0.25]{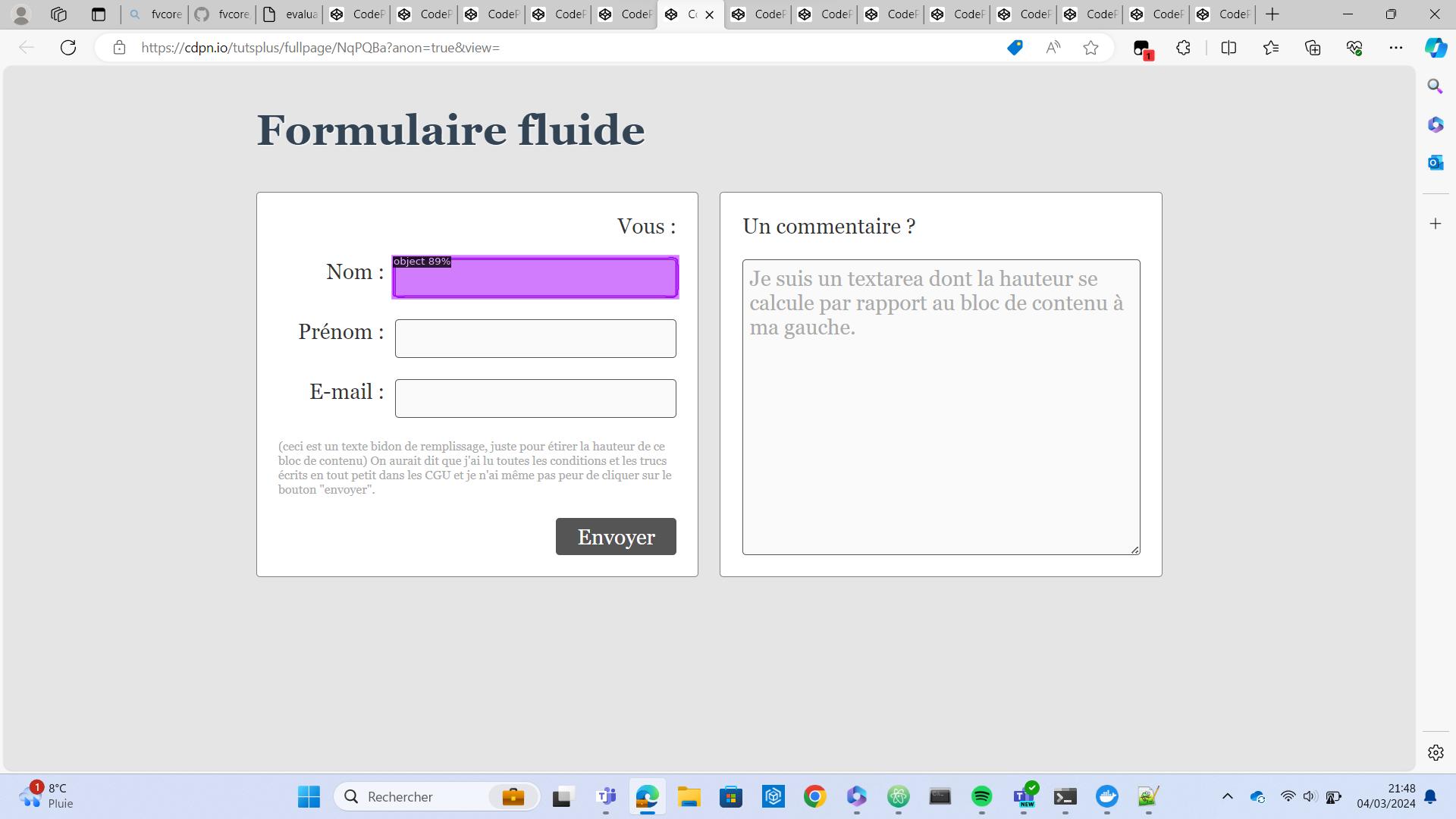}  
              \\ %\vspace{-0.75cm}  
            \end{tabular}
            \caption{Approach: IVGdirect. Reference text: \textquote{text field prénom}.}
            \label{fig:IVG_prénom}
        \end{figure}
    
    \begin{figure}[!ht]
        \centering
        \begin{tabular}{c}
          \includegraphics[scale=0.25]{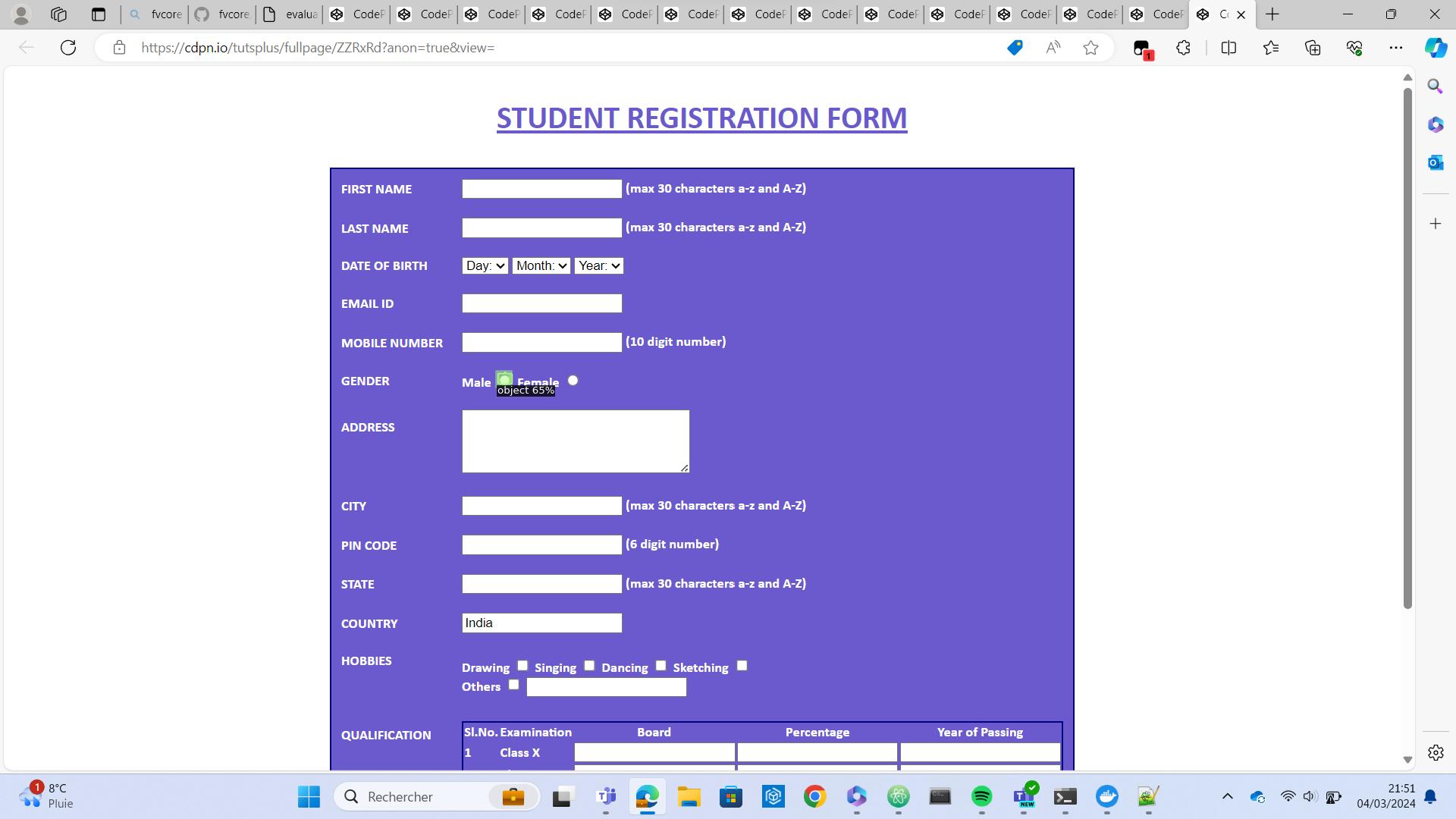}\\ %\vspace{-0.75cm}  
        \end{tabular}
        \caption{Approach: IVG. Reference text: \textquote{radio button female}.}
        \label{fig:IVGdirect_male}
    \end{figure}




%\begin{thebibliography}{00}


%%\bibitem[Lamport(1994)]{lamport94}
%%  Leslie Lamport,
%%  \textit{\LaTeX: a document preparation system},
%%  Addison Wesley, Massachusetts,
%%  2nd edition,
%%  1994.

%\end{thebibliography}

%\bibliographystyle{plain}
\bibliographystyle{elsarticle-harv} 
%\bibliography{sn-bibliography}
\bibliography{sn-bibliography}

\end{document}